\documentclass[useAMS,usenatbib,a4]{mn2e}
\usepackage[dv ips]{graphicx}
\usepackage{multirow}
\usepackage{natbib}
\usepackage{upgreek,amssymb}

\newcommand{\mnras}{MNRAS}

\newcommand{\pasp}{PASP}
\newcommand{\aap}{A\&A}

\title[The progenitor of SN iPTF13bvn and its companion]{The disappearance of the helium-giant progenitor of the type Ib supernova iPTF13bvn and constraints on its companion}  \author[J. J. Eldridge \& J.
  R. Maund]{J. J. Eldridge$^{1}$ \thanks{E-mail:
    j.eldridge@auckland.ac.nz}, J. R. Maund$^{2}$
  \\ $^{1}$Department of Physics, University of Auckland, Private Bag
  92019, Auckland, New Zealand\\ $^{2}$Department of Physics \&
  Astronomy, University of Sheffield, Hicks Building, Hounsfield Road,
  Sheffield S3 7RH, UK }

\pagerange{\pageref{firstpage}--\pageref{lastpage}} \pubyear{2005}
\begin{document}
\maketitle
\label{firstpage}

\begin{abstract}
We report and discuss post-explosion observations of supernova
iPTF13bvn. We find that the brightness of the SN at $+$740 days is
below the level of the pre-explosion source and thus confirm that the
progenitor star has gone.  We estimate that the late-time brightness
is still dominated by the supernova, which constrains the magnitude
and thus mass of a possible companion star to below approximately
10M$_{\odot}$. In turn this implies that the progenitor's initial mass
is constrained to a narrow range of between 10 to 12~M$_{\odot}$. The
progenitor of iPTF13bvn would have been a helium giant rather than a
Wolf-Rayet star. In addition, we suggest that sufficiently deep
observations acquired in 2016 would now stand a chance to directly
observe the companion star.
\end{abstract}


\section{Introduction}

\citet{Cao13} reported the detection of a possible progenitor for the
Type Ib Supernova (SN) iPTF13bvn in pre-explosion observations of NGC
5806. Due to this being the first such detected candidate for a
hydrogen-free type Ib/c SN progenitor, the pre-explosion observations
were reanalysed by a number of authors including \citet{Gro13},
\citet{Fre14} and \citet{Ber14}. The latter study by
\citeauthor{Ber14} suggested that the progenitor was probably a binary
system composed of two relatively low mass stars, in which an
interaction between the two components was responsible for removing
hydrogen from the progenitor. This model was consistent with the
prediction by \citet{yoon} that it would be easier to detect these
helium giant systems than more massive Wolf-Rayet (WR) progenitor
stars. Conversely, \citet{Gro13} suggested that a WR star progenitor
could still be possible if the progenitor had been rapidly rotating on
the main-sequence.

In \citet{Eld15} it was shown that the initial magnitudes reported by
\citet{Cao13} were in error and the pre-explosion source was brighter. The new
photometry of the pre-explosion source was compared to a large range of interacting binary
models from the BPASS (Binary Population and Spectral Synthesis)
code\footnote{\texttt{http://bpass.auckland.ac.nz}}. A number of
models were found, with initial masses from 9 to 20~M$_{\odot}$, that
matched the source in the pre-explosion images as well as the
supplementary constraints from modelling the SN, as discussed by
\citet{Fre14} and \citet{Ber14}. In these results a standard
non-rotating single-star model was strongly disfavoured as a possible
progenitor.

Here, we report on the detection of the supernova in late-time observations
at magnitudes, finding it to be fainter than the pre-explosion source and faint
enough that the mass of the companion star can be constrained. This
constraint again strongly disfavours a Wolf-Rayet star as the
progenitor of the SN and the best model progenitors are helium giants
with an initial mass of approximately 10 to 12M$_{\odot}$.

\vspace{-10pt}
\section{On the late-time supernova brightness}
\label{s2}

\begin{figure*}
\includegraphics[angle=270,width=55mm]{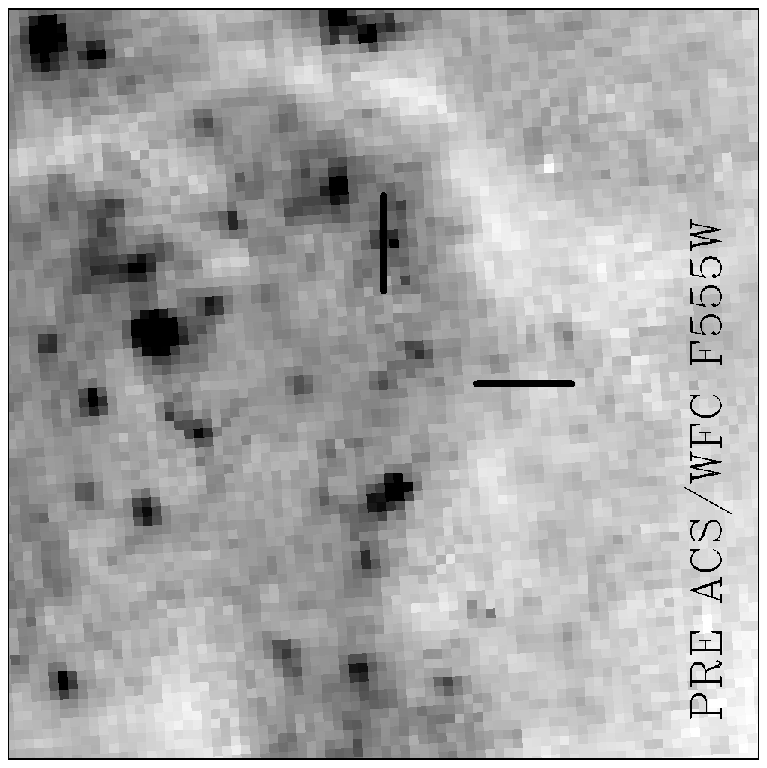}
\includegraphics[angle=270,width=55mm]{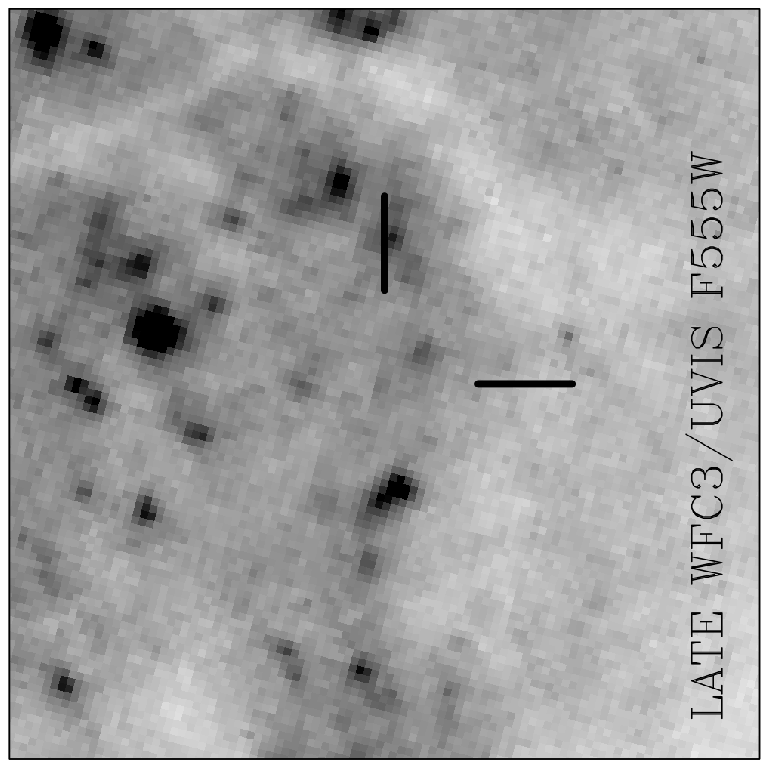}
\includegraphics[angle=270,width=55mm]{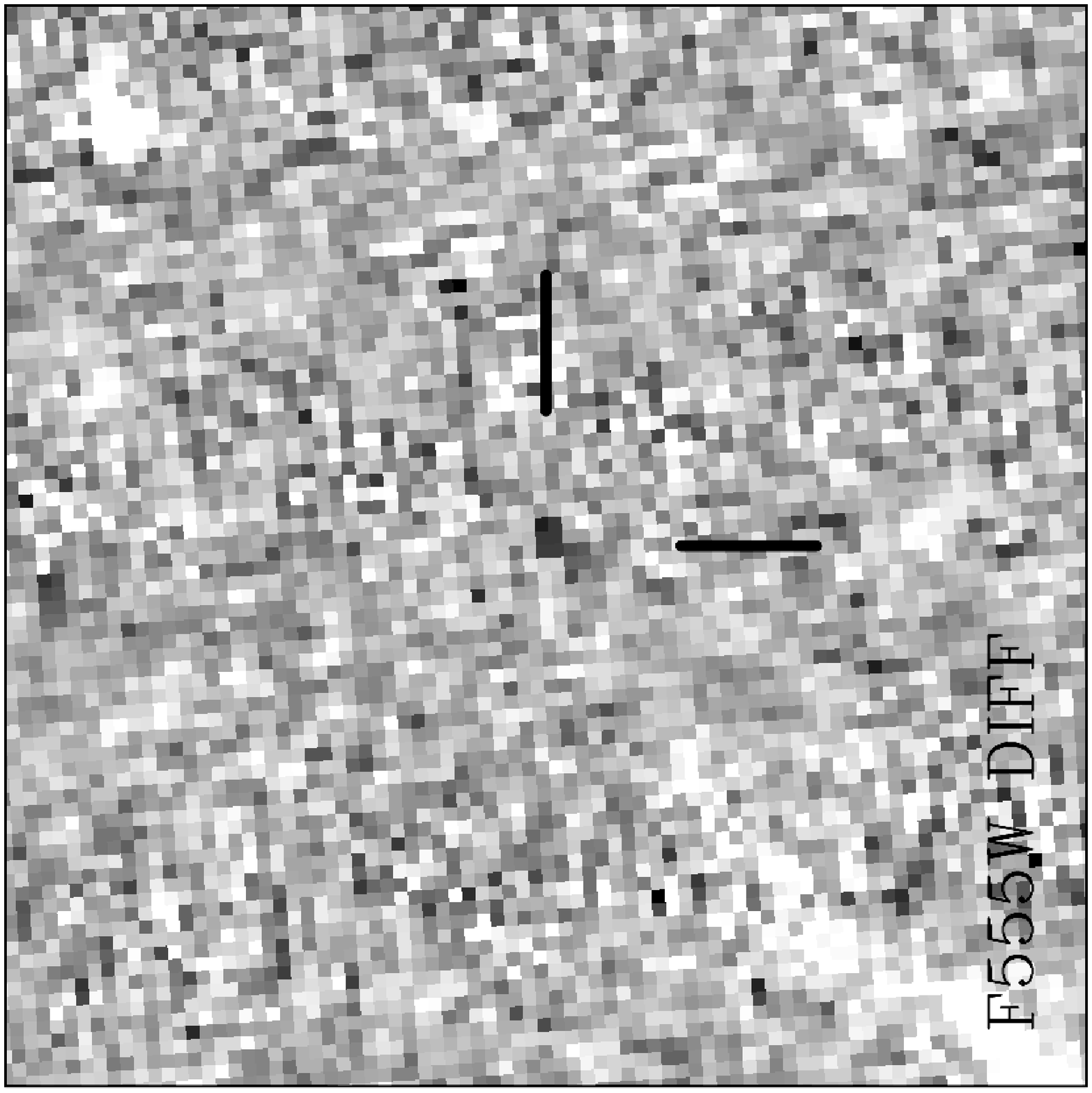}
\caption{Pre-explosion and late-time (left and centre)
  HST observations of the site of iPTF13bvn, acquired with the ACS/WFC
  and WFC3/UVIS. The SN position on all frames is indicated by the
  cross-hairs.  On the right is shown a difference image, calculated between the pre-explosion and late-time $F555W$ observations using the {\sc isis} image subtraction package \citep{Alard98,Alard00}.}
\label{fig1}
\end{figure*}

The pre-explosion observations of iPTF13bvn, in which the progenitor
candidate was identified, were previously presented by \citet{Cao13}
and \citet{Eld15}. The brightness of the progenitor object was found
to be {\it F435W} = 25.80$\pm$0.12, {\it F555W} = 25.80$\pm$0.11, {\it
  F814W} = 25.88$\pm$0.24 mags.
  
Late-time observations of the site of iPTF13bvn were acquired with
the Wide Field Camera 3 (WFC3) Ultraviolet-Visible channel (UVIS) on
26 June 2015 \citep[approximately 740 days post-explosion][]{Cao13}.
The observations, which were immediately made public, were acquired as
part of programme GO-13684 (P.I. S. Van Dyk).  The observations were
composed of two sequences of exposures of $2 \times 2860$s and
$2860+2750$s using the F438W and F555W filters, respectively.  The
constituent exposures were combined using the {\it astrodrizzle}
package, running in the PyRAF environment\footnote{STSDAS and PyRAF
  are products of the Space Telescope Science Institute, which is
  operated by AURA for NASA}.  The position of the SN was identified
on the late-time images through comparison with post-explosion WFC3
UVIS observations of iPTF13bvn acquired on 2 September 2013 as part
of programme GO-12888 (P.I. S. Van Dyk).  A geometric transformation
between the post-explosion and late-time images was calculated using
35 common stars, with resulting uncertainty on the SN position in the
late-time images of 0.36px (or 14 milliarcseconds).  A source is
recovered on the late-time images within $1\sigma$ from the
transformed SN position.  Photometry of the late-time images was
conducted using the DOLPHOT
package\footnote{http://americano.dolphinsim.com/dolphot/}
\citep{dolphhstphot} and the brightness of the source at the SN
position was measured to be $m_{F438W} = 26.48 \pm 0.08$ and
$m_{F555W} = 26.33 \pm 0.05$\, mags.  We note that the sharpness
parameter, measured by DOLPHOT, was found to be $-0.430$ and $-0.560$
in the $F438W$ and $F555W$ images, respectively.  These values are
just outside the range of sharpness values expected for point-like
sources, that may suggest that the recovered source is slightly
extended.  The SN position on the pre-explosion and late-time $F555W$ images is shown on Figure \ref{fig1}.

The post-explosion source is 0.68 and 0.53 mags fainter than the
pre-explosion source in the $F438W$ and $F555W$ observations; although
we note that the pre-explosion and late-time $B$-band observations
were acquired with slightly different filters. The brightness of the
post-explosion source is below the level of the pre-explosion source,
confirming that the pre-explosion progenitor candidate was the
progenitor which has now disappeared.

In \citet{Eld15}, we derived a range of absolute magnitudes in each
filter due to the uncertainty in the extinction towards the
pre-explosion source.  We reuse the same values here, we adopt a
distance of 22.5$\pm$2.4 Mpc, $\upmu=31.76\pm0.36$ mag and assume the
reddening is between $\mathrm{(}E(B-V)=0.045\mathrm{)}$ mag and
0.17$\pm$0.03 mag. We note that there is an updated distance from
\citet{2013AJ....146...86T}, who report a distance modulus of
$\upmu=32.14\pm 0.20$. To allow for this extra uncertainty we use this
greater distance to calculate the upper limits and the shorter
distance for the lower limits. For the {\it F435W} and {\it F555W}
observations these were between $-6.15 \geq M_{F435W} \geq -6.89$ and
$-6.1 \geq M_{F555W} \geq -6.71$. Here we find the range of absolute
magnitudes for the post-explosion source are between $-5.47 \geq
M_{F435W} \geq -6.21$ and $-5.57 \geq M_{F555W} \geq -6.18$. With
these magnitudes below the previous pre-explosion source we know that
source was the progenitor which has now disappeared. We find using the
distance does not significantly alter the nature of our fit but does
decrease the mean metallicity slightly.

To estimate whether this source could still be the SN or another
object we have compared our photometry with earlier $B$ and $V$-band
observations of \citet{Sri14} and \citet{Kun15}. We list the known B
and V band magnitudes in Table \ref{obs1}. We then use the last
observed magnitudes and the decay rate from \citet{Kun15} to predict
what the expected magnitudes should be on the date of the HST
observations. The B/F438W magnitudes are quite close and we thus can
only assume that this flux arises from the SN itself. We see however
that the V/F555W magnitudes are quite different with the observed flux
being significantly brighter than expected from the previous behaviour
of the lightcurve.  We note, however, that \citet{Kun15} measured the
$V$-band decay rate to be 1.55 magnitudes per 100 days, which is
significantly higher than the decay rates of 1.13 and 1.32 measured
for the B and R band photometry.

To gain some insight into the possible late-time behaviour of iPTF13bvn
we have also considered the similar evolution of SN 2011dh, as described
by \citet{Erg15} and Fremling et al. (in prep).  At 700 days the $B-V$
colour of 2011dh is very similar to the value we observed for iPTF13bvn. We also see
that the decay rate became shallower at 700 days. These studies also
indicate that emission lines can provide a significant contribution to
the broadband fluxes at different times. We suggest that the
$V$ band flux has been affected by nebular emission in the past so that
predicting the long term behaviour has some degree of uncertainty. Even
so it is still likely that the fluxes we observe for the late-time source are consistent with
the flux arising from the SN.

If we assume that long term evolution should be mostly similar and
that it becomes shallower over a longer time we can assume the decay
rates between the different filters should be more similar. Using our
photometry and comparing them to the latest magnitudes reported by
\citet{Sri14} we find decay rates of 1.20 and 1.30. The similarity
between the decay rates suggests that the source detected is, in fact,
the SN and not a companion star.

\begin{table}
\caption{Photometry of iPTF13bvn}
\label{obs1}
\begin{center}
\begin{tabular}{cccc}
\hline
\hline
Date   & Phase{$^\dagger$} &   $m(B/F438W)$  &  $m(V/F555W)$ \\
           & (days)                     & (mag)                & (mag) \\
\hline
06/08/2013  & 53.93 &  18.2   &   -- \\
10/09/2013  &  88.84& --     &  17.8 \\
18/04/2014  & 324 & 21.1   &  21.2  \\
Expected    & 740 & 26.0   &  28.0 \\
26/06/2015  & 740 & 26.5   &  26.3 \\
Expected   & 1001 & 29.6   & 29.7\\
\hline
\hline
\end{tabular}
$^{\dagger}$ with respect to the explosion date June 15.67 2013 \citet{Cao13}
\end{center}
\end{table}

\vspace{-10pt}
\section{Numerical Method}

\begin{figure*}
\includegraphics[angle=0, width=160mm]{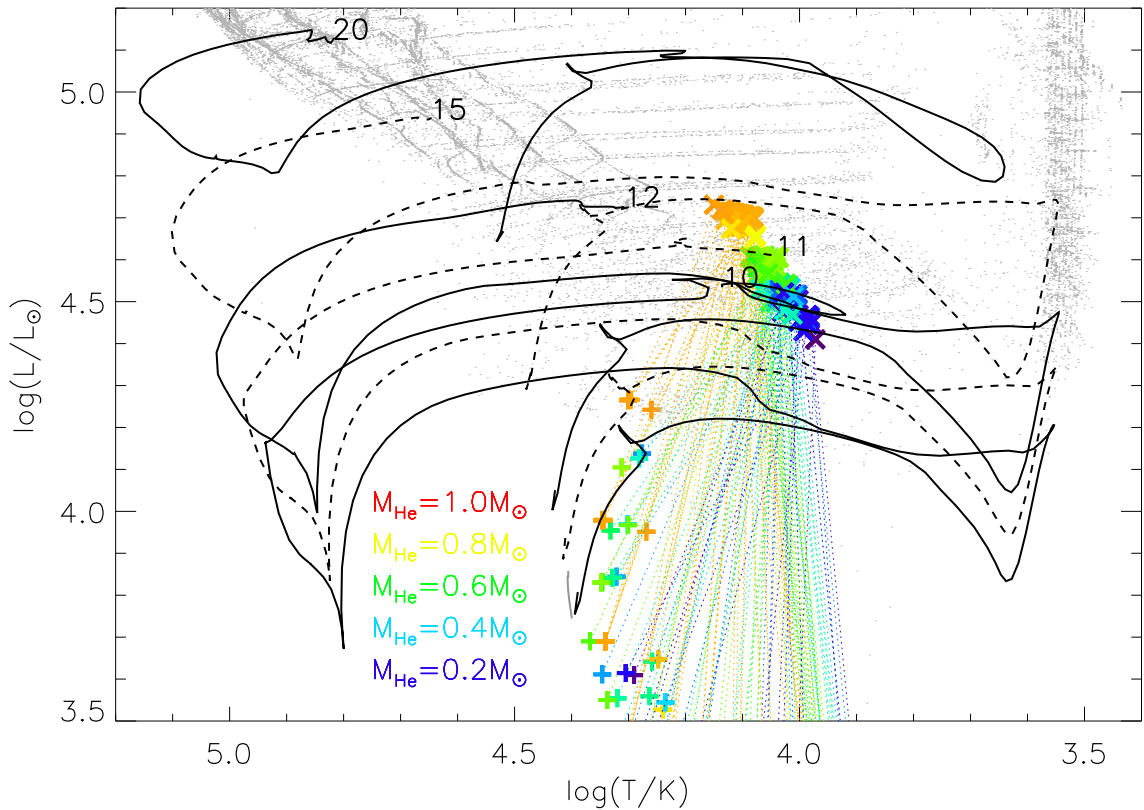}
\caption{HR diagram showing the location of possible binary progenitor
  models (primary and companions) applicable to iPTF13bvn. The '$\times$' show the location of the
  progenitor while the '$+$' is the companion. The colour indicates
  the mass of helium in the progenitor. The grey dots represent all
  possible progenitors of supernovae. The black solid and dashed lines
  are representative stellar evolution tracks for the primary
  star. The end point of each model is labelled with 
  the stars' initial mass. The tracks are ordered in increasing initial mass of 10, 11, 12, 15 and
  20M$_{\odot}$.}
\label{fig2}
\end{figure*}

In this letter we use the latest BPASS v2.0 models rather than the
previous v1.1 as used in \cite{Eld13,Eld15}. These models are almost
identical in construction as described in detail in \cite{Eld08},
however, there are more initial masses, more initial separations, more
masses for each companion star and a wider range of metallicities.
Full details of v2.0 will be described in Eldridge et al. (in
prep). They have already been used in a few studies including
\citet{SEB2016} and \citet{2016arXiv160103850W}, and have been found to match previous results well.
Here we use these models and compare them to the progenitor candidate
in a similar method to \citet{Eld15} with an extended grid of stellar
models. The key difference to the analysis of \citeauthor{Eld15} is
that we can now use the late-time HST photometry of the source at the
position of iPTF13bvn to constrain the brightness of a possible
companion.

We show the details of the resulting matching models in Figure
\ref{fig2}, in which the locations of the primary and companion stars
are shown on the Hertzsprung-Russell (HR) diagram. In Figure
\ref{fig3} we compare the pre-explosion and late-time magnitudes to
the pre- and post-explosion magnitudes expected from our stellar
models. We also provide the mean parameters for the progenitor and
companion calculated from all the models that fit the observations in
Table \ref{tab1}.

\vspace{-10pt}
\section{Results \& Discussion}
\label{s4}

\begin{table}
\caption{Physical parameters of the binary progenitor models which
  match the observed constraints on the progenitor of iPTF13bvn.}
\label{tab1}
\begin{center}
\begin{tabular}{cccc}
\hline
\hline
Primary &&Secondary&\\
Parameter & Value &Parameter & Value \\
\hline
$M_{1,i}/M_{\odot}$ & 11.0$\pm$1.2 &$M_{2,i}/M_{\odot}$ & 5.8$\pm$2.9 \\
$M_{1,f}M_{\odot}$ & 2.4$\pm$0.4  &$M_{2,f}/M_{\odot}$ & 5.0$\pm$4.5 \\
$\log(L_1/L_{\odot})$ &   4.6$\pm$0.1  &     $\log(L_2/L_{\odot})$ &   1.1$\pm$2.9     \\
$\log(T_{1, \rm eff}/K)$    &  4.06$\pm$0.04  &   $\log(T_{2, \rm eff}/K)$    &  4.0$\pm$0.4     \\
$\log(R_1/R_{\odot})$    &  1.71$\pm$0.04  &   $\log(R_2/R_{\odot})$    &  0.4$\pm$0.3     \\
$M_{\rm ejecta}M_{\odot}$ &0.95$\pm$0.4&& \\
$M_{\rm He}M_{\odot}$ & 0.6$\pm$0.2 &&\\
\hline
\hline
\multicolumn{4}{c}{System Parameters}\\
\hline
$\log(P_i/{\rm days})$ & 1.9$\pm$0.5 & $\log(a_f/R_{\odot})$   &  1.8$\pm$0.2 \\
Age/Myrs    &   24  $\pm$ 5   &  $Z$ &0.027$\pm$0.013\\
\hline
\hline
\end{tabular}
\end{center}
\end{table}

\begin{figure}
\includegraphics[angle=0, width=87mm]{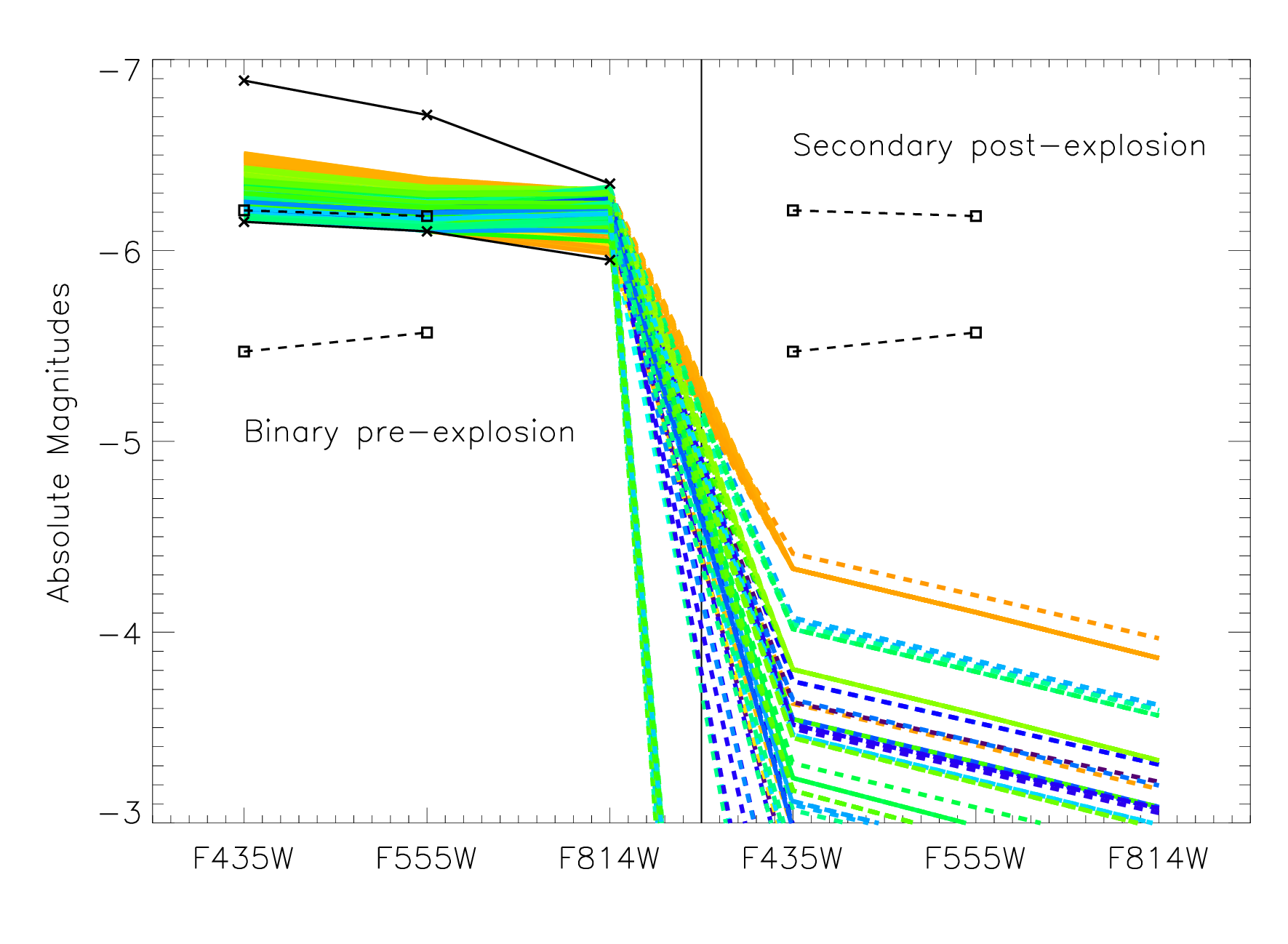}
\caption{SEDs of progenitor models compared to limits derived here with
  both the low and high extinction values used.  The observed
  magnitudes have an error of $\pm 0.3$ mags. The left hand panel
  shows the pre-explosion magnitudes for out models and the right-hand
  panel shows the post-explosion magnitudes. Pre-explosion observed
  magnitudes are shown in solid black lines while the dashed lines
  represent the observed post-explosion magnitudes. Here the colours
  of the lines represent the helium abundance of the model following the scheme used in Figure \ref{fig2}.}
\label{fig3}
\end{figure}

Figure \ref{fig3} shows that in comparison to the similar diagram in
\citet{Eld15} the preferred progenitor is now a cooler helium giant,
rather than a Wolf-Rayet star. The main reason from this constraint
can be understood in Figure \ref{fig3}. The predicted pre-explosion
magnitudes agree well with the observed magnitudes for low extinction,
while all the predicted post-explosion magnitudes are at least 1
magnitude fainter than the late-time observed photometry. The more
massive 17 to 20M$_{\odot}$ models from \citet{Eld15} are ruled out
because most of the pre-explosion magnitudes have a significant
contribution from the companion star which would be brighter than the
late-time magnitudes.

The mean parameters for the companion are shown in Table
\ref{tab1}. The one value that is relatively unconstrained is the
metallicity. While the mean metallicity is above Solar we find fits to
the progenitor between a mass fraction of $Z=0.008$ to 0.040, or
[O/H]=8.5 to 9.2, the lower range of which is equivalent to that
observed by \citet{Kun15}. We note that the radius
predicted from out models of $\sim 50R_{\odot}$ is within the range
allowed by the study of \citet{Ber14}.

The secondary mass is also constrained to around 6M$_{\odot}$, however
the allowed range is quite large. We are able to rule out a star more
massive than approximately 20M$_{\odot}$ via the late time observations.

\begin{figure}
\includegraphics[angle=0, width=85mm]{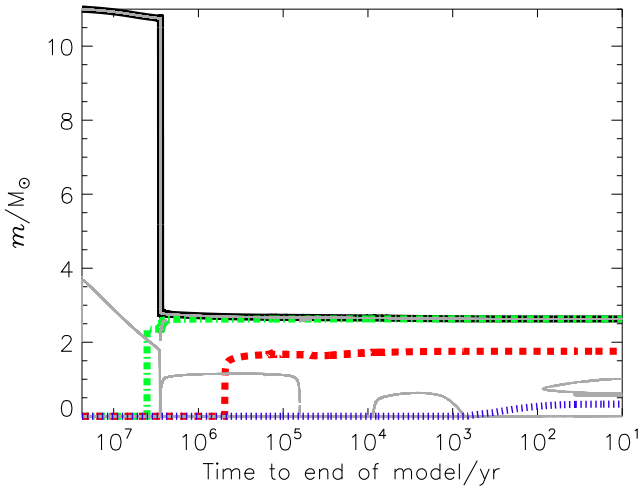}
\caption{Kippenhahn diagram for an example progenitors of iPTF13bvn in
  the upper panel. The initial mass of the binaries stars were 11 and
  5.5M$_{\odot}$ in a orbit with an initial period of 63 days. The
  thick black line represents the total stellar mass, the green
  dash-dotted line the mass of the helium core, the red dashed line
  the CO core mass and the blue dotted line the ONe core. The thin
  grey lines represent the convective boundaries.}
\label{fig-kipp}
\end{figure}

We show the evolution of an example system in Figure
\ref{fig-kipp}. The hydrogen envelope is lost during a period of
common-envelope evolution after the main sequence. We see little
further mass is lost, otherwise interior evolution progresses as it
would with the hydrogen envelope (although the helium core does not
experience dredge-up to reduce its mass).

An interesting prediction from our work is that in a sufficiently long
observation we might expect a late-time image of the supernova
location to reveal the companion as the supernova should have faded
below the possible companion magnitudes. For example, at 1000 days
after explosion the SN brightness would be of order $\sim$29.6 mags
(as shown in Table \ref{obs1}), corresponding to an absolute magnitude
of -2.1. The brightest companions we predict have an absolute
magnitude of the order of -4, an apparent magnitude of 27.7. It is
also possible, however, that a companion might be be in the form of a
black hole or neutron star, which would not be observable.

\vspace{-10pt}
\section{Conclusion}

We have shown that the progenitor star of iPTF13bvn has disappeared
and most likely had an initial mass between 10 to 12~M$_{\odot}$ and
had a companion star with a much lower mass; otherwise it would have
been observable in the post-explosion images. The mass range for the
progenitor is a factor of two less than that suggested by
\cite{Ber14}, however this was because they used the incorrect
photometry of \cite{Cao13}. We note they still suggest the progenitor
was a helium giant although the final helium star mass is
3.5M$_{\odot}$.

Our conclusions are strongly based on the caveat that the source seen
in the late-time image is still the supernova and not the companion
star and that we can use the supernova magnitude as an upper limit
for the companion star.

We note that a recent similar deep search for the companion to SN
1994I, a type Ic event, also did not find any progenitor
\citep{vandyk}. This SN is considerably closer, so observations of the
same depth reach lower absolute magnitudes of approximately -3.4 and
derived a similar mass limit for the progenitor as we find here. In
the case of iPTF13bvn we have a firm constraint on the nature of the
progenitor star and therefore the age of any companion star, this
allows us to take account of the fact that stars become brighter and
cooler as they evolve. In light of this the limit on the mass of the
companion to SN~1994I from \citet{vandyk} maybe be less than they
found. Their study and our own however does begin to suggest that if
the progenitor stars of type Ib and Ic SNe are low-mass helium giants,
the progenitors may be bright, their companions are also low mass and
most likely faint and thus difficult to observe.

Another caveat is these helium giants have still not been observed at
any other time so their observable signature is uncertain as are the
theoretical atmosphere spectra used to predict their brightness and
colours. Here the progenitor is too cool for using the Potsdam
Wolf-Rayet spectra so we use the BaSeL spectra. The problem here is
that those models are hydrogen rich. \citet{yoonrecent} have
calculated possible spectra using the detailed atmosphere modelling
code CMFGEN; however, one problem is that these giants have low
surface gravity which can significantly alter the shape of the
spectrum. We confirm that we cannot fit \textit{any} Wolf-Rayet star
models to the pre-explosion photometry, for iPTF13bvn, and still have
a companion below the late-time magnitudes, whilst having an ejecta
mass below 3M$_{\odot}$.

We also wish to make the community aware of the opportunity that there
is now a better chance of detecting the companion star as the
supernova should now have faded to the point where the companion
should be directly detectable if a sufficently long observation with
HST was taken.

\section{Acknowledgements}

The authors thank the referee Christoffer Fremling for positive
comments that greatly improved the paper. JJE also thanks
Morgan Fraser and Stephen Smartt for comments on an early draft of this
letter. JJE acknowledges support from the University of Auckland. The
research of JRM is supported by a Royal Society Research
Fellowship. The authors wish to acknowledge the contribution of the
NeSI high-performance computing facilities and the staff at the Centre
for eResearch at the University of Auckland. New Zealands national
facilities are provided by the New Zealand eScience Infrastructure
(NeSI) and funded jointly by NeSIs collaborator institutions and
through the Ministry of Business, Innovation and Employments
Infrastructure programme. URL: http://www.nesi.org.nz.nz.



\label{lastpage}
\bsp

\end{document}